\documentstyle[
amsmath,amssymb]{article}

\begin{document}

\oddsidemargin 5mm

\newtheorem{lem}{Lemma}[section]
\newtheorem{th}{Theorem}[section]
\newtheorem{prop}{Proposition}[section]
\newtheorem{rem}{Remark}[section]
\newtheorem{define}{Definition}[section]
\newtheorem{cor}{Corollary}[section]

\allowdisplaybreaks

\makeatletter\@addtoreset{equation}{section}\makeatother
\def\theequation{\arabic{section}.\arabic{equation}}

\newcommand{\D}{{\cal D}}

\newcommand{\N}{{\Bbb N}}
\newcommand{\C}{{\Bbb C}}
\newcommand{\Z}{{\Bbb Z}}
\newcommand{\R}{{\Bbb R}}
\newcommand{\Rp}{{\R_+}}
\newcommand{\eps}{\varepsilon}
\newcommand{\om}{\omega}
\newcommand{\hot}{\hat\otimes}

\newcommand{\supp}{\operatorname{supp}}
\newcommand{\la}{\langle}
\newcommand{\ra}{\rangle}
\newcommand{\const}{\operatorname{const}}
\newcommand{\FF}{{\cal F}}
\newcommand{\rom}[1]{{\rm #1}}
\renewcommand{\author}[1]{\medskip{\large #1}\par\medskip}

\newcommand{\tob}{\Subset}
\newcommand{\dd}{\overset{{.}{.}}}

\newcommand{\rhoa}{\rho_{\mathrm a}}

\renewcommand{\emptyset}{\varnothing}

\newcommand{\per}{\operatorname{per}}
\newcommand{\indlim}{\operatornamewithlimits{ind\,lim}}

\renewcommand{\kappa}{\varkappa}

\newcommand{\lw}{{:}\,}
\newcommand{\rw}{\,{:}}

\begin{center}{\Large \bf
Fermion and boson random point processes as particle distributions
of infinite free  Fermi and Bose gases of finite density
}\end{center}

\author{EUGENE LYTVYNOV}

\noindent{\sl Institut f\"{u}r Angewandte Mathematik,
Universit\"{a}t Bonn, Wegelerstr.~6, D-53115 Bonn, Germany;
\newline SFB 256, Univ.~Bonn, Germany;
\newline BiBoS, Univ.\ Bielefeld, Germany}

\begin{abstract}
\noindent The aim of this paper is to show that fermion and boson
random point processes naturally appear from representations of
CAR and CCR which correspond to gauge invariant generalized free
states (also called quasi-free states). We consider particle
density operators $\rho(x)$, $x\in\R^d$, in the representation of
CAR describing an infinite free Fermi gas of finite density at
both zero and finite temperature  \cite{AW}, and in the
representation of CCR describing an infinite free Bose gas at
finite temperature \cite{ArWoods}. We prove that the spectral
measure  of the smeared operators $\rho(f)=\int dx\, f(x)\rho(x)$
(i.e., the measure $\mu$ which allows to realize the $\rho(f)$'s
as multiplication operators by $\la\cdot,f\ra$ in $L^2(d\mu)$) is
a well-known fermion, resp.\ boson process on the space of all
locally finite configurations in  $\R^d$.
\end{abstract}

\section{Introduction: representations of current algebras; boson and fermion processes}
The nonrelativistic quantum mechanics of many identical particles
may be described by means of a field $\psi(x)$, $x\in\R^d$,
satisfying either canonical commutation relations (CCR) and
describing bosons: \begin{gather} [\psi(x),\psi(y)]_- =
[\psi^*(x),\psi^*(y)]_-=\pmb0 ,\notag\\ [\psi^*(x),\psi(y)]_-
=\delta(x-y)\pmb1\label{z7eawr76},\end{gather}
 or satisfying
canonical anticommutation relations (CAR) and describing fermions:
\begin{gather} [\psi(x),\psi(y)]_+ =
[\psi^*(x),\psi^*(y)]_+ =\pmb0,\notag\\ [\psi^*(x),\psi(y)]_+
=\delta(x-y)\pmb1.\label{tzew45}\end{gather} Here,
$[A,B]_\mp=AB\mp BA$ is the commutator, resp.\ anticommutator. The
statistics of the system is thus determined by the algebra which
is to be represented.

In the formulation of nonrelativistic quantum mechanics in terms
of particle densities and currents, one defines
\begin{align}\rho(x)&{:=}\psi^*(x)\psi(x),\notag\\
J(x)&{:=}(2i)^{-1}\big( \psi^*(x)\nabla
\psi(x)-(\nabla\psi^*(x))\psi(x)\big).\label{u8zearg6}\end{align}
Using CCR or CAR, one can formally compute the commutation
relations satisfied by the smeared operators
$\rho(f){:=}\int_{\R^d}dx\, f(x)\rho(x)$ and
$J(v){:=}\int_{\R^d}dx\, v(x)\cdot J(x)$. These turn out to be
\begin{gather}[\rho(f_1),\rho(f_2)]_-=\pmb0,\notag\\
[\rho(f),J(v)]_-=i\rho(v\cdot\nabla f),\notag\\
[J(v_1),J(v_2)]_-=-iJ(v_1\cdot\nabla v_2-v_2\cdot \nabla
v_1),\label{zue6ed}\end{gather}
 independently of whether
one starts with CCR or CAR.

Thus, in a nonrelativistic current theory, the particle statistics
is not determined by a choice of an equal-time algebra, but
instead may be determined by the choice  of a representation of
the algebra, see e.g.\ \cite{DaSh,Go,GoMeSha}  and the references
therein. Since the operators $\rho(f)$ and $J(v)$ are generally
speaking unbounded, one usually starts with study of the group $G$
obtained by exponentiating the algebra $\frak g$ generated by the
commutation relations  \eqref{zue6ed}. More precisely, considering
$\rho(f)$ and $J(v)$ to be self-adjoint, the corresponding
one-parameter groups are \begin{equation} {\cal
U}(tf)=\exp\big[it\rho(f)\big],\qquad {\cal
V}(\phi_t^v)=\exp\big[it J(v))\big],\label{zftft}\end{equation}
where $\phi_t^v$ is the one-parameter group of diffeomorphisms (or
flows) on $\R^d$ generated by the vector field $v$: $$
\frac{\partial \phi_t^v}{\partial t}=v(\phi_t^v),\qquad
\phi_{t=0}^v(x)=x.$$ From equation \eqref{zue6ed}, the operators
\eqref{zftft} satisfy the group law $$ {\cal U}(f_1){\cal
V}(\psi_1){\cal U}(f_2){\cal V}(\psi_2)={\cal U}(f_1+f_2\circ
\psi_1){\cal V}(\psi_2\circ\psi_1).$$ Hence, the  group $G$ is the
semidirect product $S(\R^d)\wedge \operatorname{Diff}(\R^d)$,
where $S(\R^d)$ is the Schwartz space of rapidly decreasing
functions on $\R^d$ and $\operatorname{Diff}(\R^d)$ is a certain
group of diffeomorphisms of $\R^d$ of Schwartz's type (and thus
containing all diffeomorphisms with compact support, i.e., which
are identical outside a compact set). Due to physical
interpretation, the currents $\rho(f)$ and $J(v)$ (and the group
$G$) can be taken as fundamental structures of quantum mechanics.
It should be noted that, given a representation of the group $G$
and its current algebra $\frak g$, corresponding operators
$\psi(x), \psi^*(x)$ may, in general, not exist.

Let ${\cal U}(f)$ be a continuous unitary cyclic representation of
$S(\R^d)$ in a Hilbert space $\cal H$ with cyclic vector $\Omega$.
The functional $L(f){:=}({\cal U}(f)\Omega,\Omega)$ satisfies the
conditions of the Bochner--Minlos theorem, and hence is the
Fourier transform of a probability measure $\mu$ on $S'(\R^d)$,
the dual of $S(\R^d)$: $$
L(f)=\int_{S'(\R^d)}\exp\big[i\la\omega,f\ra\big]\,
\mu(d\omega).$$ Therefore, ${\cal H}$ can be realized as
$L^2(S'(\R^d);\mu)$, $\Omega= 1$, and ${\cal U}(f)$ as the
multiplication operator by $\exp\big[i\la\cdot,f\ra\big]$ in
$L^2(S'(\R^d);\mu)$.

Let now ${\cal U}(f){\cal V}(\psi)$ be a continuous unitary cyclic
representation of the group $S(\R^d)\wedge
\operatorname{Diff}(\R^d)$ in $\cal H$. Suppose additionally that
the cyclic vector $\Omega$ is also cyclic for (the smaller family
of unitary operators) ${\cal U}(f)$. From some physical reasons,
one believes  that, in the spinless case, the latter condition is
always fulfilled as long as one does not deal with parastatistics.
Then, realizing the Hilbert space ${\cal H}$ as
$L^2(S'(\R^d);\mu)$ just as above, by  \cite{Go}, one has that the
measure $\mu$ is quasi-invariant for $\operatorname{Diff}(\R^d)$
and the operators ${\cal V}(\psi)$ become
\begin{equation}\label{zgufgtgf12} ({\cal V}(\psi)F)(\omega)=
\chi_\psi(\omega)F(\psi^*\omega)\bigg(\frac{d\mu(\psi^*\omega)}{d\mu(\omega)}\bigg)^{1/2},
\qquad \text{$\mu$-a.e.\ $\omega\in S'(\R^d)$},\end{equation} for
all $F\in L^2(S'(\R^d);\mu)$. Here,
$d\mu(\psi^*\omega)/d\mu(\omega)$ is the Radon--Nikodym derivative
of the transformed measure with respect to the original measure,
and $\chi_\psi(\omega)$ is a cocycle, i.e., $\chi_\psi(\cdot)$ is
a complex-valued function of modulus one, depending on $\psi$,
defined $\mu$-a.e., and satisfying, for each
$\psi_1,\psi_2\in\operatorname{Diff}(\R^d)$,
$\chi_{\psi_2}(\omega)\chi_{\psi_1}(\psi_2^*\omega)=\chi_{\psi_1\circ\psi_2}(\omega)$
$\mu$-a.e.

A powerful method  of construction of continuous unitary cyclic
representations of $G$ has been the method of generating
functional. A continuous complex-valued function $E$ on $G$ is
called a generating functional on $G$ if the following conditions
are fulfilled: 1) $E(e)=1$, where $e$ is the identity element of
$G$; 2) $\sum_{i,j=1}^N\bar\lambda_i\lambda_j E(g_i^{-1}g_j)\ge0$
for all $\lambda_i\in\C$, $g_i\in G$, $i=1,\dots,N$, $N\in\N$. By
Araki's theorem \cite{Arrr}, $E$ is a generating functional on $G$
if and only if there exists a continuous unitary cyclic
representation $\pi$ of $G$ in $\cal H$ with cyclic vector
$\Omega$ such that $E(g)=(\Omega,E(g)\Omega)$, $g\in G$. Thus, one
may implicitly construct unitary representations of $G$ by finding
generating functionals.

In \cite{GoGrPoSha} (see also \cite{AKR1, AKR2,VGG}),  the case of
an infinite free Bose gas at zero temperature with  average
particle density $\rho>0$ was studied in the formalism of local
current algebras. Goldin {\it et al}.\ started with considering a
system of $N$ bosons in a box of volume $V$. The physical Hilbert
space $\cal H$ is now $L^2_{\mathrm s}(V^N)$, the subspace of
$L^2(V^N)$ consisting of all symmetric functions (we also used the
letter $V$ to denote the box itself). The Hamiltonian for $N$
boson particles is given by $H_{N,V}=-\frac12\sum_{i=1}^N\Delta_i$
with periodic boundary conditions, and the normalized ground state
wave function $\Omega_{N,V}=V^{-N/2}$. The representation of the
group $G$ in the box $V$ is given by \begin{align} {\cal
U}_{N,V}(f)F(x_1,\dots,x_N&)=\exp\bigg(i\sum_{i=1}^N
f(x_i)\bigg)F(x_1,\dots,x_N),\notag\\ {\cal
V}_{N,V}(\psi)F(x_1,\dots,x_N)&=F(\psi(x_1),\dots,\psi(x_N))\prod_{i=1}^N
\sqrt{J_\psi(x_i)},\label{574}\end{align} where $f$ and $\psi$
have support inside the box $V$ and $J_\psi(x)=\det(\partial
\psi^k(x)/\partial x_l)_{k,l=1}^d$ is the Jacobian of the flow.
Thus, one can write down the generating functional
\begin{equation}\label{gzdrtdtds} E_{N,V}(f,\psi)=
(\Omega_{N,V},{\cal U}_{N,V}(f){\cal
V}_{N,V}(\psi)\Omega_{N,V})\end{equation} of this representation
and take the so-called $N/V$-limit, i.e., the limit as
$N,V\to\infty$, $N/V\to\rho$. The limiting functional then has the
form $$
E(f,\psi)=\exp\bigg(\rho\int_{\R^d}\big(e^{if(x)}\sqrt{J_\psi(x)}-1\big)\,dx\bigg).$$
The authors then showed that $\Omega$ is cyclic for ${\cal U}(f)$,
$f\in S(\R^d)$, and hence this representation can be realized on
the space $L^2(S'(\R^d);\mu)$, where the Fourier transform of the
measure $\mu$ is equal to
\begin{equation}\label{zugqzgu}\int_{S'(\R^d)}\exp\big(i\la\omega,
f\ra\big)\,\mu(d\omega)=\exp\bigg(\rho\int_{\R^d}\big(e^{if(x)}-1\big)\,dx\bigg).\end{equation}
Thus, $\mu=\pi_\rho$ is the Poisson measure with intensity $\rho\,
dx$. This measure is concentrated on the space $\Gamma_{\R^d}$ of
all locally finite configurations in $\R^d$. As for the operators
${\cal V}(\psi)$ in this representation, the general formula
\eqref{zgufgtgf12}  now takes  the following form: all the
cocycles are identically equal to one and the Radon--Nykodym
derivative is given by $$\frac
{d\pi_\rho(\psi^*\gamma)}{d\pi_\rho(\gamma)}=\prod_{x\in\gamma}J_\psi(x),\qquad
\text{$\pi_\rho$-a.e.\ $\gamma\in\Gamma_{\R^d }.$}$$ One may also
derive an explicit formula for the action of the operators $J(v)$
in $L^2(\Gamma_{\R^d};\pi_\rho)$, which particularly shows that
these are certain differential operators on the configuration
space (see \cite{AKR1} for details).

Furthermore, it was shown in \cite{GoGrPoSha} that the
representation of $G$ defined by \eqref{gzdrtdtds} is  unitarily
equivalent to the representation in the symmetric Fock space
$\FF_{\mathrm s}(L^2(\R^d))$ in which the operators $\rho(x)$,
$J(x)$ are defined by formula \eqref{u8zearg6} with
\begin{equation}\label{uilsagfz} \psi(x)=\psi_{\mathrm
F}(x)+\sqrt\rho,\qquad \psi^*(x)=\psi^*_{\mathrm F}(x)+\sqrt\rho
\end{equation}  and $\Omega$ is the vacuum vector in $\FF_{\mathrm s}(L^2(\R^d))$.
In \eqref{uilsagfz}, $\psi_{\mathrm F}(x)$, $\psi^*_{\mathrm
F}(x)$ are the standard annihilation and creation operators in the
Fock space, respectively. In fact, this unitary equivalence has
played a crucial role in the study of the representation defined
by \eqref{gzdrtdtds}.

On the other hand, the obtained unitary $I:{\cal F}_{\mathrm
s}(L^2(\R^d))\to L^2(\Gamma_{\R^d};\pi_\rho)$ coincides with the
well-known chaos decomposition  for the Poisson measure (e.g.\
\cite{Sur}). The operators
$$\rho(x)=\psi^*(x)\psi(x)=\psi_{\mathrm F}^*(x)\psi_{\mathrm
F}(x)+\sqrt\rho\,\psi^*_{\mathrm F}(x)+\sqrt{\rho}\,\psi_{\mathrm
F}(x)+\rho$$ are known in quantum probability as quantum Poisson
white noise (e.g.\ \cite{HuPa,mey}).
 The $\pi_\rho$ can also be  thought of as the spectral
measure of the family $(\rho(f))_{f\in S(\R^d) }$ (cf.\
\cite[Ch.~3]{BeKo} and  \cite{Ber,Ly}).

In fact, it was in Araki and Woods' paper \cite{ArWoods} dealing
with representations of CCR that the operators \eqref{uilsagfz}
first appeared in the description of an infinite free Bose gas at
zero temperature.

 In \cite{M1,M2} (see also \cite{Girard}), a unitary cyclic representation of the group $G$
describing an infinite free Fermi gas at zero temperature was, in
particular, studied. Menikoff started with a system of $N$ free
Fermi particles in a cubic box $V$ in $\R^3$ with edges of length
$L$. If $k_f$ is the Fermi momentum of the system, the number of
the particles in the box $V$ is equal to
\begin{equation}\label{guzftzf} N=\# \big\{k_n =(2\pi/L)(n_1,n_2,n_3): \, n_1,n_2,n_3\in\Z,\,
|k_n|\le k_f\big\}.
\end{equation}
The Hilbert space of the system is then ${\cal H}=L^2_{\mathrm a}
(V^N)$, the subspace of $L^2(V^N)$ consisting of all antisymmetric
functions, and the Hamiltonian of the system
$H_{N,V}=-\frac12\sum_{i=1}^N\Delta_i$. The normalized ground
state of $H_{N,V}$ is $$
\Omega_{N,V}(x_1,\dots,x_N)=(V^{-N}/N!)^{1/2}\,\det\big(\exp(ik_n\cdot
x_m)\big)_{n,m=1}^N,$$ where the $k_n$'s are as in
\eqref{guzftzf}. The representation of the group $G$  in the box
$V$ is given by the same formulas \eqref{574} but with $F\in
L^2_{\mathrm s}(V^N)$. By \eqref{guzftzf}, we get for the average
particle density $$  N/V= N/(L^3)\to \rho=(4/3)
\pi(k_f/2\pi)^3\quad\text{as }L\to\infty.$$ Taking the limit
$N,V\to\infty$, $N/V\to \rho$  of the generating functional
$E_{N,V}(f,\psi)$ again given by \eqref{gzdrtdtds}, one gets the
following results. Let $$
\kappa(x){:=}(2\pi)^{-3}\int_{\{|\lambda|< k_f\} }e^{i\lambda
x}\,d\lambda=3\rho(\sin z-z\cos z)/(z^3)\big|_{z=k_f|x|},\qquad
x\in\R^3,$$ and let $$ R_n(y_1,\dots,y_n;x_1,\dots,x_n){:=}\det
(\kappa(x_i-y_j))_{i,j=1}^n,\qquad n\in\N.$$ Then, the limiting
generating functional is given by \begin{multline}\label{56}
E(f,\psi)=1+\sum_{n=1}^\infty \int dx_1\int dy_1\dotsm \int
dx_n\int dy_n
\prod_{i=1}^n\big[\delta(x_i-y_i)(e^{if(x_i)}T_{x_i}(\psi)-1)\big]\\
\times R_n(y_1,\dots,y_n;x_1,\dots,x_n),
\end{multline} where $ T_x(\psi)
g(x){:=}g(\psi(x))\sqrt{J_\psi(x)}$. In particular,
\begin{equation}\label{jhifztfd} L(f)=1+\sum_{n=1}^\infty
\int_{(\R^3)^n}\bigg(\prod_{i=1}^n (e^{if(x_i)}-1)\bigg)
\det(\kappa(x_i-x_j))_{i,j=1}^n\, dx_1\dotsm dx_n,\end{equation}
which is the Fourier transform of a measure $\mu_{\mathrm a}$ on
$S'(\R^3)$. Furthermore, it follows from \eqref{jhifztfd} that the
measure $\mu_{\mathrm a}$ has correlation functions
\begin{equation}\label{tzdds} k_{\mu_{\mathrm a}}^{(n)}(x_1,\dots,x_n)=\det
(\kappa(x_i-x_j))_{i,j=1}^n.\end{equation} Menikoff mentioned in
\cite{M2}  that from the existence of correlation functions it
should follow that the measure $\mu_{\mathrm a}$ is concentrated
on $\Gamma_{\R^d}$, however he could not prove it.

Two important problems remained open after \cite{M1,M2}: 1) Is
there any connection with the representation of CAR for an
infinite free Fermi gas at zero temperature? Does corresponding
$\psi(x),\, \psi^*(x)$ operators exist? 2) Is $\Omega$---the
cyclic vector of the representation defined by \eqref{56}---also
cyclic for the ${\cal U}(f)$'s, and if it is so, what is the form
of the Radon--Nikodym derivative and the cocycles in the formula
\eqref{zgufgtgf12} in this case?

When studying statistical properties of a chaotic beam of fermions
by using wavepacket formalism, Benard and Macchi \cite{BeMa}
arrived at measures on the configuration space over a bounded
volume whose correlation functions are given by a formula of type
\eqref{tzdds}. In \cite{Ma1,Ma2}, Macchi called a measure on the
configuration space a fermion process if the respective
correlation functions are given by \eqref{tzdds} in which
$\kappa(x-y)$ is a non-negative definite function. She gave
sufficient conditions for the existence of such a measure. Fermion
process (also called determinantal random point fields) are often
met with in random matrix theory, probability theory,
representation theory, and ergodic theory. We refer  to the paper
\cite{Sosh} containing an exposition of recent as well as
sufficiently old results on the subject. Scaling limits of fermion
point processes are proved in \cite{Scale1,Scale2,Spohn}. We also
refer to  the recent papers \cite{Shi, ShiYoo} for a discussion of
different problems connected with fermion processes on the
configuration space  over the lattice $\Z^d$.

In a parallel way, a boson process was defined and studied in
\cite{BeMa,Ma0,Ma1}. This process is defined as a probability
measure $\mu$ on $\Gamma_{\R^d}$ whose correlation functions are
given by $$k_\mu^{(n)}=\per(\kappa(x_i-x_j))_{i,j=1}^n,$$ where
$\kappa(x-y)$ is again a non-negative definite function and the
permanent $\per A$ of a matrix $A$ contains the same terms as the
corresponding  determinant $\det A$ but with constant positive
signs for each product of matrix elements in place of the
alternating  positive and negative signs of the determinant. It
should be mentioned that any boson process is a  Cox process,
i.e., a Poisson process with a random intensity measure, see
\cite[Sec.~8.5]{DV}.

In \cite{F,FF} (see also \cite{FF2}), point processes were
constructed and studied which correspond to locally normal states
of a boson system, and which can be interpreted as the position
distribution of the state. More precisely, let ${\cal L}({\cal F
}_{\mathrm s}(L^2(\R^d)))$ be the von Neumann algebra of all
bounded linear operators in the symmetric Fock space ${\cal F
}_{\mathrm s}(L^2(\R^d))$, and let $\cal A$ be its
$C^*$-subalgebra obtained as the uniform closure of all local von
Neumann algebras in  ${\cal L}({\cal F }_{\mathrm s}(L^2(\R^d)))$.
Let $\omega$ be a locally normal state on $\cal A$ (cf.\
\cite{BraRo}). As well known, the symmetric Fock space ${\cal F
}_{\mathrm s}(L^2(\R^d))$ may be isomorphically realized as the
$L^2$-space $L^2(\Gamma_{\R^d}^{\mathrm fin};\lambda)$, where
$\Gamma_{\R^d}^{\mathrm fin}$ is the space of all finite
configurations in $\R^d$ and $\lambda$ is the Lebesgue--Poisson
measure on $\Gamma_{\R^d}^{\mathrm fin}$. (Notice the evident
inclusion $\Gamma_{\R^d}^{\mathrm fin}\subset \Gamma_{\R^d}$.)
Then, every bounded function $F$ on $\Gamma_{\R^d}^{\mathrm fin}$
determines a bounded operator ${\cal M}_F$ of multiplication by
$F$. A function $F$ on $\Gamma_{\R^d}$ is called local if there
exists a compact set $\Lambda\subset \R^d$ such that
$F(\gamma)=F(\gamma\cap\Lambda)$ for all $\gamma\in\Gamma_{\R^d}$.
The restriction of $F$ to $\Gamma_{\R^d}^{\mathrm fin}$ is
 a local function on $\Gamma_{\R^d}^{\mathrm fin}$, for
which we preserve the notation $F$. In \cite{FF}, it was proved
that there exists a unique probability measure $\mu_\omega$ on
$\Gamma_{\R^d}$ such that, for all bounded local functions $F$ on
$\Gamma_{\R^d}$,
$$\int_{\Gamma_{\R^d}}F(\gamma)\,\mu_\omega(d\gamma)=\omega({\cal
M }_F).$$ Some properties of such point processes were also
studied in \cite{FF}. In \cite{FF2}, it was shown that, if the
reduced density matrices
$\rho^{(n)}_\omega(x_1,\dots,x_n;y_1\dots,y_n)$ of the state
$\omega$ exist and are continuous, then the correlation functions
of the measure $\mu_\omega$ are given by \begin{equation}
k_{\mu_\omega}^{(n)}(x_1,\dots,x_n)=\rho_\omega^{(n)}(x_
1,\dots,x_n;x_1,\dots,x_n).\label{zggtfft}\end{equation}
Furthermore, the special case corresponding to the ideal Bose gas
(cf.\ \cite{BraRo}) was studied in detail in \cite{F}. By
\eqref{zggtfft},  $\mu_\omega$ is now the boson measure with $$
\kappa(x)=\sum_{n=1}^\infty \frac{z^n}{(4\pi \beta
n)^{d/2}}\,\exp\big[-|x|^2/(4n\beta)\big],$$ where $\beta$ is the
inverse temperature and $z$ the activity. This boson measure was
proved to be an infinite divisible point process. It should also
be noted that the local normality of the states we discuss below,
in Section~2, was established in e.g.\ \cite{BraRo}.

The aim of this paper is to show a connection between
representations  of CAR, resp.\ CCR  describing infinite free
Fermi, resp.\ Bose gases of finite density and the fermion, resp.\
boson random point processes.

In Section~2, we recall Araki and Wyss' representations of CAR in
the the antisymmetric Fock space that  describes  infinite free
Fermi gases at both finite and zero temperature \cite{AW}, and
Araki and Woods' representation of CCR in the symmetric Fock space
that describes an infinite free Bose gas at finite temperature
\cite{ArWoods}. The  results of this section are essentially known
(with the only exception that the corresponding ``annihilation''
and ``creation'' operators $\psi(x),\psi^*(x)$ has not been
treated without smearing).

In Section~3, we prove that the corresponding particle density
operators are well-defined and form a family $(\rho(f))_{f\in
S(\R^d)}$ of commuting selfadjoint operators. Then, we introduce
the space ${\frak H}_{\sharp}$ ($\sharp=$a in the fermionic case
and $\sharp=$s in the bosonic case) as the closed linear span of
the vectors of the form $\rho(f_1)\dotsm\rho(f_n)\Omega$.
Restricted to this space, $(\rho(f))_{f\in S(\R^d)}$ evidently
becomes a cyclic family. Using the spectral theory of cyclic
families of commuting selfadjoint operators \cite{BeKo,Sam}, we
then show that there exist a unique probability measure
$\mu_\sharp$ on $S'(\R^d)$---the Schwartz space of tempered
distributions---and a unitary operator $I_\sharp:{\frak
H}_\sharp\to L^2(S'(\R^d);\mu_\sharp)$ such that
$I_\sharp\Omega=1$ and
$I_\sharp\rho(f)I_\sharp^{-1}=\la\cdot,f\ra\cdot$ for each $f\in
S(\R^d)$.

Next, we introduce an operator field
$\lw\rho(x_1)\dotsm\rho(x_n)\rw$ via a recurrence relation, and
prove that $$\lw\rho(x_1)\dotsm\rho(x_n)\rw=\psi^*(x_n)\dots
\psi^*(x_1)\psi(x_1)\dotsm\psi(x_n).$$  Thus,
$\lw\rho(x_1)\dotsm\rho(x_n)\rw$ is nothing but a normal (Wick)
product of $\rho(x_1),\dots,\rho(x_n)$. Using this and results of
\cite{BKKL}, we explicitly calculate the correlation functions of
$\mu_\sharp$. This enables us, first, to show that $\mu_\sharp$ is
concentrated on the configuration space $\Gamma_{\R^ d}$, and
second, to identify $\mu_{\mathrm a}$ as a fermion process, and
$\mu_{\mathrm s}$ as a boson process. In particular, starting from
the representation of CAR \cite{AW} corresponding to an infinite
free Fermi gas at zero temperature, we arrive at the the same
probability measure $\mu_{\mathrm a}$ as Menikoff did in
\cite{M2}.

Thus, the main results of the paper are as follows: 1) We
introduced  the particle density operators (quantum  white noise)
$\rho(x)$, $x\in\R^d$, corresponding to an infinite free Fermi gas
(of finite density) at zero and at finite temperature, resp.\ an
infinite free Bose gas at finite temperature, proved the
well-definedness of $\rho(x)$ and  the essential self-adjointness
of the corresponding field $(\rho(f))_{f\in S(\R^d)}$. 2) We
proved a one-to-one correspondence between the operator field
$(\rho(f))_{f\in S(\R^d)}$ and a fermion, resp.\ boson point
process $\mu_\sharp$ on $\Gamma_{\R^d}$. Furthermore, the
constructed unitary isomorphism between the spaces ${\frak
H}_{\sharp}$ and $L^2(\Gamma_{\R^d};\mu_{\sharp})$ can be thought
of as a kind of a chaos decomposition of
$L^2(\Gamma_{\R^d};\mu_{\sharp})$ (compare with the Poisson case).

We also note that, though the very existence of a fermion process
under a slightly stronger condition on the function $\kappa$ in
terms of its Fourier transform has been known before (cf.\
\cite[Proposition~4.1]{BO}), as a by-product of our results we get
a new proof of the existence of fermion (as well as boson)
processes.

It is still an open problem to show that also the operators $J(v)$
may be realized on $L^2(\Gamma_{\R^d};\mu_\sharp)$, but we hope
that the obtained  unitary operator between the latter space and
the corresponding subspace of the Fock space may be of some help
to tackle this problem.

\section{Infinite free Fermi and Bose gases of finite density}\label{gzdtese}

 We first recall the construction of a cyclic representation of
 CAR whose state (constructed with respect to the cyclic
vector) is a gauge invariant generalized free state. This
representation  is due to Araki and Wyss \cite{AW}. Generalized
free states were first defined and studied by Shale and
Stinespring \cite{ShSt}. Since that generalized free states (also
called quasi-free states) have been studied  by several authors,
see e.g.\ \cite{Araki,BV,BraRo,DA,MV,PS,R} and the references
therein.

Let $H$ be a separable real Hilbert space and let $H_{\C}$ denote
its complexification.  We suppose that the scalar product in
$H_{\C}$, denoted by $(\cdot,\cdot)_{H_{\C}}$,  is antilinear in
the first dot and linear in the second one. Let $$\FF_{\mathrm a}
(H)\big(=\FF_{\mathrm a}(H_\C)\big){:=}\bigoplus_{n=0}^\infty
\FF_{\mathrm a}^{(n)} (H)$$ denote the antisymmetric Fock space
over $H$. Here, $\FF_{\mathrm a}^{(0)}(H){:=}\C$ and, for
$n\in\N$, $\FF_{\mathrm a}^{(n)}(H){:=}H_{\C}^{\wedge n}$,
$\wedge$ standing for antisymmetric tensor product. By
$\FF_{\mathrm a,\, fin}(H)$ we denote the subset of $\FF_{\mathrm
a}(H)$ consisting of all elements $f=(f^{(n)})_{n=0}^\infty\in
\FF_{\mathrm a}(H)$ for which $f^{(n)}=0$, $n\ge N$, for some
$N\in \N$. We endow $\FF_{\mathrm a,\, fin}(H)$ with the topology
of the topological direct sum of the spaces $\FF_{\mathrm a}^{(n)}
(H)$. Thus, the convergence in $\FF_{\mathrm a,\, fin}(H)$ means
uniform finiteness and coordinate-wise convergence.

For $f\in H_\C$, we denote by $a(f)$ and $a^*(f)$ the standard
annihilation and creation operators on $\FF_{\mathrm a}(H)$. They
are defined on the domain $\FF_{\mathrm a,\, fin}(H)$ through the
formula
\begin{align} a(f) h_1\wedge\dotsm\wedge h_n &{:=}\frac1{\sqrt
n}\, \sum_{i=1}^n (-1)^{i+1} (f,h_i)_{H_{\C}} h_1\wedge \dotsm
\wedge h_{i-1}\wedge \check h_i\wedge h_{i+1}\dotsm \wedge
h_n,\notag \\ a^*(f) h_1\wedge\dotsm\wedge h_n &{:=}\sqrt{n+1}\,
f\wedge h_1\wedge\dotsm\wedge h_n,\label{ftwsea}
\end{align}
where $h_1,\dotsm,h_n\in H_{\C}$ and  $\check h_i$ denotes the
absence of $h_i$. The operator $a^*(f)$ is the restriction to
$\FF_{\mathrm a,\, fin}(H)$ of the adjoint of $a(f)$ in
$\FF_{\mathrm a}(H)$, and both  $a(f)$ and $a^*(f)$ act
continuously on $\FF_{\mathrm a,\, fin}(H)$. The annihilation and
creation operators satisfy  CAR:
\begin{gather*}[a(f),a(g)]_+=\pmb 0,\\ [a(g),a^*(f)]_+
=(g,f)_{H_{\C}} \pmb 1 \end{gather*} for all $f,g\in H_{\C}$.

Let $K$ be a linear operator in $H_\C$ such that $\pmb 0\le K\le
\pmb 1$. We take the direct sum $H\oplus H$ of two copies of the
Hilbert space $H$, and construct the antisymmetric Fock space
$\FF_{\mathrm a}(H\oplus H)$. For $f\in H_\C$, we denote
$a_1(f){:=}a(f,0)$, $a_2(f){:=}a(0,f)$ and analogously $a_i^*(f)$,
$i=1,2$. Let also $K_1{:=}K^{1/2}$, $K_2{:=}(\pmb 1-K)^{1/2}$. We
then set, for $f\in H_\C$,
\begin{equation} \psi(f){:=} a_2 (K_2f)+a^*_1(JK_1f),\qquad \psi^*(f){:=}
 a_2^* (K_2f)+a_1(JK_1f),\label{gztde45}\end{equation} where $J:H_\C\to H_\C$ is the
 operator of complex conjugation: $Jf{:=}\overline f$.
 As easily
seen, the operators $\{\psi(f),\psi^*(f)\mid f\in H_\C\}$ again
satisfy  CAR. Let $H_i$ denote the closure  of
$\operatorname{Im}K_i$ in $H_\C$, $i=1,2$. Then, restricted to the
subspace $\FF_{\mathrm a}(H_1\oplus H_2)$, the operators
$\{\psi(f),\psi^*(f)\mid f\in H_\C\}$ form a cyclic representation
of
 CAR with cyclic vector $\Omega{:=}(1,0,0,\dots)$---the vacuum
in $\FF_{\mathrm a}(H\oplus H )$.

Let ${\frak A}_{\mathrm a}(H_\C)$ denote the $C^*$-algebra
generated by the operators $\psi(f)$, $f\in H_\C$, and let
$\omega_{\mathrm a}$ be the state on ${\frak A}_{\mathrm a}(H_\C)$
defined by $\omega_{\mathrm
a}(\Psi){:=}(\Psi\Omega,\Omega)_{\FF_{\mathrm a}(H_1\oplus H_2)}$,
$\Psi\in{\frak A}_{\mathrm a}(H_\C)$. The $n$-point functions of
$\omega_{\mathrm a}$ are given by the formula
\begin{equation}\label{32rsfd} \omega_{\mathrm a}(\psi^*(f_n)\dotsm \psi^*(f_1)\psi(g_1)\dotsm
\psi(g_m))=\delta_{n,m} \det ((f_i,Kg_j)_{H_\C})\end{equation} for
all $f_1,\dots,f_n,g_1,\dots, g_m\in H_\C$. Thus, $\omega_{\mathrm
a}$ is a gauge invariant generalized free state corresponding to
the operator $K$.

An analogous representation of  CCR was constructed by Araki and
Woods  \cite{ArWoods} (historically  it preceded the
representation of  CAR \cite{AW}). Let us outline it. In the
symmetric Fock space  over $H$, denoted by $\FF_{\mathrm s}(H)$,
we  construct the standard annihilation and creation operators,
$b(f)$ and $b^*(f)$, which satisfy  CCR:
\begin{gather*}[b(f),b(g)]_- =\pmb0,\\ [b(g),b^*(f)]_- =(g,f)_{H_\C}\pmb 1
\end{gather*} for all $f,g\in H_\C$. Let now
${\cal K}$ be a bounded linear operator in $H_\C$ such that ${\cal
K}\ge\pmb0$. We set ${\cal K}_1{:=}{\cal K}^{1/2}$, ${\cal
K}_2{:=}(\pmb1+{\cal K})^{1/2}$. Analogously to \eqref{gztde45},
we define the following operators in $\FF_{\mathrm s}(H\oplus H)$:
\begin{equation} \varphi(f){:=} b_2
({\cal K}_2f)+b_1^*(J{\cal K}_1f),\qquad \varphi^*(f){:=}b_2^*
({\cal K }_2 f)+b_1(J{\cal K}_1 f)\label{gztfdrde45}\end{equation}
for $f\in H_\C$. These operators again satisfy  CCR and form a
cyclic representation of  CCR in the Hilbert space $\FF_{\mathrm s
}({\cal H}_1\oplus {\cal H}_2)$, where ${\cal H}_i$ is the closure
of $\operatorname{Im}{\cal K}_i$ in $H_\C$, $i=1,2$. Let ${\frak
A}_{\mathrm s}(H_\C)$ denote the $C^*$-algebra generated by the
operators $\varphi(f)$, $f\in H_\C$, and let $\omega_{\mathrm s}$
be the state on ${\frak A}_{\mathrm s}(H_\C)$ defined by $
\omega_{\mathrm s}(\Psi){:=}(\Psi\Omega,\Omega)_{\FF_{\mathrm
s}(H\oplus H)}$, $\Psi \in {\frak A}_{\mathrm s}(H_\C)$. The
$n$-point functions of  $\omega_{\mathrm s}$  are  given by
 \begin{equation}\label{gtfr56e54}
 \omega_{\mathrm s}(\varphi^*(f_n)\dotsm \varphi^*(f_1)\varphi(g_1)\dotsm
\varphi(g_m))=\delta_{n,m} \per((f_i,Kg_j)_{H_\C}).\end{equation}

We now proceed to consider an infinite free Fermi gas of finite
density, which is a special case of representation
\eqref{gztde45}. Let $H{:=}L^2(\R^d;dx)=L^2(\R^d)$, and so
$H_{\mathrm\C}=L^2_{\C}(\R^d)=L^2(\R^d\to\C;dx)$. To fix
notations, we define the Fourier transform of a function $f\in
L_{\C}^1(\R^d)$ by $$ \FF f(\lambda){:=}\hat f(\lambda) {:=}
(2\pi)^{-d/2}\int_{\R^d} e^{-i x\cdot \lambda} f(x)\,dx,\qquad
\lambda \in\R^d,$$ and the inverse Fourier transform by
$$\FF^{-1}f(x){:=}\check f(x){:=}(2\pi)^{-d/2} \int_{\R^d}
e^{i\lambda\cdot x}f(\lambda)\,d\lambda,\qquad x\in\R^d,$$ so that
$\FF$ can be extended by continuity from $L_{\C}^2(\R^d)\cap
L_{\C}^1(\R^d)$ to a unitary operator on $L_{\C}^2(\R^d)$, and
$\FF^{-1}$ is the inverse operator of $\FF$.

Let $k$ be the inverse Fourier transform of a function $\hat k$
satisfying the following conditions:
\begin{equation}\label{rde4w3}0\le\hat k\le1,\quad \hat k\in
L^1(\R^d).\end{equation} We define $K{:=}\FF^{-1}\hat k\cdot\FF$,
where $f\cdot$ denotes the operator of multiplication by a
function $f$. Using this $K$, we construct the operators
$\psi(f),\psi^*(f)$ defined in the Fock space $\FF_{\mathrm
a}(H\oplus H)$ by formula \eqref{gztde45}. Notice that now
$$H_{1}=\FF^{-1}L_{\C}^2(\operatorname{supp}\hat k;dx),\qquad
H_{2}=\FF^{-1}L_{\C}^2(\operatorname{supp}(1-\hat k);dx).$$

This  representation of CAR  describes an infinite free Fermi gas
with density distribution $\hat k(\cdot)$ in ``momentum space''
\cite{AW}, see also \cite{DA}. In particular, if
 $\beta$ is the inverse
temperature, $\mu$ the chemical potential, and $m$ the mass of a
particle, the corresponding infinite free Fermi gas is described
by
\begin{equation}\label{gzter}\hat
k(\lambda)=\frac{\exp(\beta\mu-\beta\,\frac{|\lambda|^2}{2m})}{1+\exp(\beta\mu-\beta\,\frac{|\lambda|^2}{2m})},\qquad
\lambda\in\R^d.\end{equation}  For the limit $\beta\to\infty$ of
zero temperature, we obtain
\begin{equation}\label{gzftrsres}\hat
k(\lambda)=\pmb1_{B(\sqrt{2m\mu})}(\lambda),\qquad
\lambda\in\R^d,\end{equation} where $B(r)$ denotes the ball in
$\R^d$ of radius $r>0$ centered at the origin, and
$\pmb1_X(\cdot)$ denotes the indicator of a set $X$. Notice that,
in the case of \eqref{gzter}, we have $H_{1}=H_{2}=H_{\C}$, while
in the case of \eqref{gzftrsres} $H_{\C}=H_{1}\oplus H_{2}$.

We will now need a rigging of $\FF_{\mathrm a}(H)$. Let ${\cal
D}(\R^d)$ denote the space  of all real-valued infinite
differentiable functions on $\R^d$ with compact support. For
$p\in\N$, we define a weighted Sobolev space $S_p(\R^d)$ as the
closure of ${\cal D }(\R^d)$ with respect to the Hilbert norm $$
\|f\|_p^2{:=}\int_{\R^d}A^p f(x)f(x)\,dx,\qquad f\in{\cal
D}(\R^d),$$ where
\begin{equation}\label{z76t45}Af(x){:=}-\Delta
f(x)+(|x|^2+1)f(x),\qquad x\in\R^d,\end{equation} is the harmonic
oscillator. We identify $S_0(\R^d)=L^2(\R^d)$ with its dual and
obtain $$ S(\R^d){:=}\projlim_{p\to\infty}S_p(\R^d)\subset
L^2(\R^d)\subset \indlim_{p\to\infty}S_{-p}(\R^d){=:} S'(\R^d).$$
We recall that the Fourier transform $\FF$ is a continuous
bijection of $S_{\C}(\R^d)$ onto $S_{\C}(\R^d)$, and, extended by
continuity, it is a continuous bijection of $S_{\C}'(\R^d)$ onto
$S_{\C}'(\R^d)$. Here, $S_\C(\R^d)$ and  $S'_\C(\R^d)$ denote the
complexification of $S(\R^d)$ and  $S'(\R^d)$, respectively.

Denoting $\Phi{:=}S_{\C}(\R^d)\oplus S_{\C}(\R^d) $,
$\Phi_p{:=}S_{p,\C}(\R^d)\oplus S_{p,\C}(\R^d)$, and
$\Phi'{:=}S'_{\C}(\R^d)\oplus S'_{\C}(\R^d)$, we  get $
\Phi=\projlim_{p\to\infty}\Phi_p$ and $\Phi'=
\indlim_{p\to\infty}\Phi_{-p}$. We set, for $n\in\Z_+$,
$$\FF_{\mathrm a}^{(n)}(\Phi){:=}\projlim_{p\to\infty}\FF_{\mathrm
a}^{(n)}(\Phi_p),\qquad \FF_{\mathrm
a}^{(n)}(\Phi'){:=}\indlim_{p\to\infty}\FF^{(n)}_{\mathrm
a}(\Phi_{-p}).$$ Let $\FF_{\mathrm a,\, fin} (\Phi)$ denote the
topological direct sum of the spaces $\FF_{\mathrm
a}^{(n)}(\Phi)$, $n\in\Z_+$. 
The dual of
$\FF_{\mathrm a,\, fin (\Phi)}$ with respect to the zero space
$\FF_{\mathrm a}(H\oplus H )$ is $\FF_{\mathrm a,\,
fin}^*(\Phi)=\times_{n=0}^\infty \FF^{(n)}_{\mathrm a}(\Phi')$,
the topological product of the spaces $\FF^{(n)}_{\mathrm
a}(\Phi')$. It consists of all sequences of the form
$F=(F^{(0)},F^{(1)},F^{(2)},\dots)$ such that $F^{(n)}\in
\FF^{(n)}_{\mathrm a}(\Phi')$, and  convergence in $\FF_{\mathrm
a,\, fin}^*(\Phi)$ means  coordinatewise convergence. Thus, we
have constructed the nuclear triple $$\FF_{\mathrm a,\,
fin}(\Phi)\subset \FF_{\mathrm a}(H \oplus H)\subset \FF_{\mathrm
a,\, fin}^*(\Phi).$$

Noting that $\hat k^{1/2} \in L^2(\R^d)$ and $(1-\hat k)^{1/2}\in
S'(\R^d)$, we define $$ \kappa_1{:=} (2\pi)^{-d/2}\FF^{-1}\hat
k^{1/2}\in L_\C^2(\R^d),\qquad
\kappa_2{:=}(2\pi)^{-d/2}\FF^{-1}(1-\hat k)^{1/2}\in
S_\C'(\R^d).$$ Then, for any $f\in L_\C^2(\R^d)$, $$
K_1f(x)=\kappa_1*f(x)=\int_{\R^d}\kappa_1(x-y) f(y)\,dy,\qquad
\text{a.e.\ $x\in\R^d$},$$ and for any $f\in S_\C(\R^d)$,
$K_2f=\kappa_2* f$, where the convolution of  a generalized
function with a test one is defined in the usual way. For each
$x\in \R^d$, we define $\kappa_{1,x}\in L_\C^2(\R^d)$ and
$\kappa_{2,x}\in S_\C'(\R^d)$ by \begin{equation}\label{new}\la
\kappa_{i,x},f\ra=\la\kappa_i,f(x+\cdot)\ra,\qquad f\in
S_\C(\R^d),\ i=1,2,\end{equation}
 where
$\la\cdot,\cdot\ra$ denotes the dual pairing (generated by the
scalar product in $H_\C$). Then, for any $f\in S_\C(\R^d)$,
\begin{equation}K_i f(x)=\la \kappa_{i,x},f\ra,\qquad\text{a.e.\
$x\in\R^d$},\ i=1,2.\label{sw5e45w}\end{equation}

Using  formulas \eqref{ftwsea}, we can easily define, for each
$(f_1,f_2)\in\Phi'$, an annihilation operator $a(f_1,f_2)$ acting
continuously on $\FF_{\mathrm a,\,fin}(\Phi) $, and a creation
operator $a^*(f_1,f_2)$ acting continuously on $\FF^*_{\mathrm
a,\,fin}(\Phi) $. Analogously to the above, we then introduce
operators $a_i(f)$ and $a^*_i(f)$, $i=1,2$, for each $f\in
S_\C'(\R^d)$.

Taking to notice \eqref {new} and \eqref{sw5e45w}, we now set, for
each $x\in\R^d$, $$\psi(x){:=}a_2(\kappa_{2,x})+a^*_1(\overline
\kappa_{1,x}),\qquad
\psi^*(x){:=}a^*_2(\kappa_{2,x})+a_1(\overline\kappa_{1,x}).$$
These operators act continuously from $\FF_{\mathrm a,\,fin}(\Phi)
$ into $\FF^*_{\mathrm a,\,fin}(\Phi) $, and we have the following
integral representation: for each (real-valued) $f\in S(\R^d)$
\begin{equation}\label{3425w4278} \psi(f)=\int_{\R^d} dx\, f(x)\psi(x),\qquad
\psi^*(f)=\int_{\R^d} dx\, f(x)\psi^*(x). \end{equation} The
integration in \eqref{3425w4278} and below is to be understood in
the following sense: for example, the first equality in
\eqref{3425w4278} means: $\la \psi(f)G_1,G_2\ra=\int_{\R^d}
f(x)\la \psi(x)G_1,G_2\ra\, dx$ for any $G_1,G_2\in \FF_{\mathrm
a,\,fin}(\Phi) $. The  operators $\psi(x),\psi^*(x)$ satisfy the
CAR \eqref{tzew45}, the formulas making sense after integration
with test functions.

Now, let us briefly consider the bosonic case. Let
$H{:=}L^2(\R^d)$ and let $k$ be the inverse Fourier transform of a
function $\hat k$ satisfying the following conditions:
\begin{equation}\label{763457z} 0\le \hat k\le C\quad\text{for some
}C\in(0,\infty),\quad \hat k\in L^1(\R^d).\end{equation} We define
${\cal K}{:=}\FF^{-1}\hat k\cdot\FF$, and using this $\cal K$, we
construct the operators $\varphi(f),\varphi^*(f)$ defined on the
symmetric Fock space $\FF_{\mathrm s}(H\oplus H)$ by formula
\eqref{gztfdrde45}. If we additionally suppose that $ \hat k(x)>0$
a.e.\ $x\in \R^d$, then the obtained representation of CCR
describes an infinite free Bose gas at finite temperature and with
density distribution $\hat k$ in ``momentum space''
\cite{ArWoods}, see also \cite{DA}.

Analogously to the above, we construct the triple $$\FF_{\mathrm
s,\, fin}(\Phi)\subset \FF_{\mathrm s}(H \oplus H)\subset
\FF_{\mathrm s,\, fin}^*(\Phi),$$ and using it, we make sense of
the operators $\varphi(x)$, $\varphi^*(x)$, $x\in\R^d$. These
satisfy the CCR \eqref{z7eawr76} with $\psi$ replaced by
$\varphi$.

\section{Particle density operators and their spectral measure}

We will again consider the fermionic case in detail, and then
outline the bosonic case.

\subsection{Fermionic case}

We suppose that \eqref{rde4w3} holds. For each $x\in\R^d$, we
define a particle density operator $$
\rhoa(x){:=}\psi^*(x)\psi(x).$$ Since $\overline\kappa_{1,x}\in
L^2(\R^d)$, the operator $\psi(x)$ acts continuously from
$\FF_{\mathrm fin}(\Phi)$ into $\FF_{\mathrm fin}(H\oplus H)$, and
$\psi^*(x)$ acts continuously from $\FF_{\mathrm fin}(H\oplus H)$
into $\FF^*_{\mathrm fin}(\Phi)$. Therefore,  $\rho_{\mathrm
a}(x)$ is a well-defined, continuous operator from $\FF_{\mathrm
fin}(\Phi)$ into $\FF^*_{\mathrm fin}(\Phi)$.

We then define $$\rhoa(f){:=}\int_{\R^d}dx\, f(x)\rhoa(x),\qquad
f\in S(\R^d).$$

\begin{lem}\label{74dzt} For each $f\in S(\R^d)$\rom, the operator $\rhoa(f)$ is
well-defined and continuous on\linebreak $\FF_{\mathrm
a,\,fin}(H\oplus H) $\rom.\end{lem}

\noindent {\it Proof}. 1. We first prove the statement for
$\rho_{\mathrm a,\,1}(f){:=}\int_{\R^d}dx\,
f(x)a^*_2(\kappa_{2,x})a^*_1(\overline\kappa_{1,x})$. As easily
seen, it suffices to show that
\begin{equation}\label{fze5e}\int_{\R^d}dx\,
 f(x)\kappa_{2,x}\otimes \overline\kappa_{1,x}\in H^{\otimes2}
,\end{equation} where $\otimes$ denotes the usual tensor product.
For any $g,h\in S_\C(\R^d)$, we have \begin{align}& \left\la
\int_{\R^d}dx\, f(x)\kappa_{2,x}\otimes
\overline\kappa_{1,x},g\otimes h\right\ra \notag\\ &\qquad{:=}
\int _{\R^d} f(x)\la \kappa_{2,x},g\ra\,\la
\overline\kappa_{1,x},h\ra\, dx\notag\\ &\qquad=
\int_{\R^d}f(x)(K_2g)(x)(JK_1Jh)(x)\, dx\notag
\\ &\qquad =\int_{\R^d}\overline {f(x)(K_1Jh)(x)}\, (K_2g)(x)  \,dx\notag\\
& \qquad =\int_{\R^d}\overline{(\FF(f\cdot K_1J h ))(\lambda)}
\,(\FF K_2g)(\lambda) \,d\lambda\notag\\ &\qquad =\int_{\R^d}
(2\pi)^{-d/2} \overline{\hat f * (\hat k^{1/2}\cdot\widehat
Jh)(\lambda)}(1-\hat k)^{1/2}(\lambda)\hat
g(\lambda)\,d\lambda.\notag\\ &\qquad
=\int_{\R^d}(2\pi)^{-d/2}\int_{\R^d}\overline{\hat
k^{1/2}(\xi)\widehat{Jh}(\xi)\hat f(\lambda-\xi)}\,d\xi\, (1-\hat
k )^{1/2}(\lambda)\hat g(\lambda)\,d\lambda\notag\\
&\qquad=\int_{\R^{2d}}(2\pi)^{-d/2} \overline{\hat f(\lambda+\xi)}
 (1-\hat k)^{1/2}(\lambda)\hat k^{1/2}(-\xi) \hat g(\lambda)\hat h
 (\xi)\,d\lambda\,d\xi.\label{weqew3}
\end{align}
Since $|1-\hat k|\le 1$ and $\hat k^{1/2},\hat f\in L^2(\R^d)$,
the function $$ G_f(\lambda,\xi){:=} (2\pi)^{-d/2} \hat
f(\lambda+\xi) (1-\hat k)^{1/2}(\lambda)\hat k^{1/2}(-\xi),\qquad
\xi,\lambda\in\R^d,$$ belongs to $L^2(\R^{2d})$. Therefore, by
\eqref{weqew3}, $$\left\la \int_{\R^d}dx\, f(x)\kappa_{2,x}\otimes
\overline\kappa_{1,x},g\otimes h\right\ra= \la
\FF_{2d}^{-1}(G_f),g\otimes h\ra,\qquad g,h\in S(\R^d),$$ where
$\FF_{2d}$ denotes the Fourier transform on $L_{\C}^2(\R^{2d})$.
By linearity and continuity, this implies $$\int_{\R^d}dx\,
f(x)\kappa_{2,x}\otimes
\overline\kappa_{1,x}=\FF_{2d}^{-1}(G_f)\in L^2(\R^d)^{\otimes 2}
.$$

2. We now prove the statement for $$\rho_{\mathrm a,\,
2}(f){:=}\int_{\R^d}dx\,
f(x)a^*_2(\kappa_{2,x})a_2(\kappa_{2,x}).$$ For any $g_i,h_i\in
S_{\C}(\R^d)$, $i=1,2$, we have \begin{align*} &\left\la
\int_{\R^d} dx\, f(x)a^*_2(\kappa_{2,x})
a_2(\kappa_{2,x})(g_1,g_2),(h_1,h_2)\right\ra\\ &\qquad=
\int_{\R^d}f(x)\overline{K_2g_2(x)}\, K_2h_2(x)= \la K_2(f\cdot
K_2g_2),h_2\ra,\end{align*} and therefore $$\int_{\R^d}
dx\,a^*_2(\kappa_{2,x})
a_2(\kappa_{2,x})\restriction\FF^{(1)}(\Phi)=\pmb0\oplus
\big(K_2(f\cdot K_2)\big){=:}{\cal A}_{2,f}.$$ Evidently ${\cal
A}_{2,f} $ is continuous on $H_{\C}\oplus H_{\C}$.

For any linear continuous operator $\cal A$ on $H_{\C}\oplus
H_{\C}$, we define the second quantization of $\cal A$, denoted by
$d\Gamma({\cal A })$, as the operator in $\FF_{\mathrm a}(H\oplus
H )$ with domain $D(d\Gamma({\cal A})){:=}\FF_{\mathrm a,\,
fin}(H\oplus H)$, given by $$ d\Gamma({\cal A})\restriction
\FF_{\mathrm a}^{(n)}(H\oplus H){:=}{\cal
A}\otimes\pmb1\otimes\dots\otimes\pmb1+\pmb1\otimes{\cal
A}\otimes\pmb1\otimes\dots\otimes\pmb1+\pmb1\otimes \dots\otimes
\pmb1\otimes{\cal A}.$$ $d\Gamma({\cal A})$ acts continuously on
$\FF_{\mathrm a,\, fin}(H\oplus H)$. Then, an easy calculation
shows that $\rho_{\mathrm a,\,2 }(f)=d\Gamma({\cal A}_{2,f})$.

3. Analogously, we get \begin{align*}\rho_{\mathrm a,\, 3
}(f){:=}&\int_{\R^d}dx\, f(x)
a_1(\overline\kappa_{1,x})a^*_1(\overline\kappa_{1,x})\\ =&
\int_{\R^d} f(x)\la\overline\kappa_{1,x},\overline\kappa_{1,x}\ra
\, dx\, \pmb1-d\Gamma({\cal A}_{1,f})\\=&\int_{\R^d}f(x)\,
dx\,(2\pi)^{-d}\int_{\R^d}\hat k(\lambda)\,d\lambda\,
\pmb1-d\Gamma({\cal A}_{1,f}),\end{align*} where ${\cal
A}_{1,f}{:=}\big(JK_1J(f\cdot JK_1J)\big)\oplus\pmb0$.

4. Finally, since $\rho_{\mathrm a,1}(f)$ is a continuous operator
from $\FF_{\mathrm a}^{(n)}(H\oplus H)$ into $\FF_{\mathrm
a}^{(n+2)}(H\oplus H)$ for each $n\in\Z_+$, and since
$$\rho_{\mathrm a,\, 4}(f){:=}\int_{\R^d}dx\,
f(x)a_1(\overline\kappa_{1,x})a_2(\kappa_{2,x})$$ is its adjoint,
we have that $\rho_{\mathrm a,\, 4}(f)$ acts continuously  from
$\FF_{\mathrm a}^{(n+2)}(H\oplus H)$ into $\FF_{\mathrm
a}^{(n)}(H\oplus H)$, and hence, continuously on $\FF_{\mathrm
a,\, fin}(H\oplus H)$.\quad $\blacksquare$\vspace{2mm}

Using  anticommutation relations \eqref{tzew45}, we easily prove
the following

\begin{lem}For each $f_1,f_2\in S(\R^d)$, we have on $\FF_{\mathrm a,\, fin}(H\oplus
H)$\rom:
\begin{equation}\label{hjfcz}\rhoa(f_1)\rhoa(f_2)=\rhoa(f_2)\rhoa(f_1).\end{equation}
\end{lem}

We define a Hilbert space ${\frak H}_{\mathrm a}$ as the closure
 of the linear span of the set
$$\big\{\, \Omega, \rhoa(f_1)\dotsm\rhoa(f_n)\Omega\mid
f_1,\dots,f_n\in S(\R^d),\ n\in\N\,\}$$ in $\FF_{\mathrm
a}(H\oplus H)$.

\begin{rem}\rom{It is not hard to see that  ${\frak H}_{\mathrm a}$ is a subspace
of the space
\begin{equation}\label{guhdesrw}\bigoplus_{n=0}^\infty
P_{{\mathrm a},\, 2n}\big(H_{1,\,\C}^{\otimes n }\otimes
H_{2,\,\C}^{\otimes n}\big),\end{equation} where $P_{{\mathrm
a},\, 2n}:(H_{\C}\oplus H_{\C})^{\otimes 2n}\to (H_{\C}\oplus
H_{\C})^{\wedge 2n}$ is the antisymmetrization operator.
Evidently,  \eqref{guhdesrw} is a subspace of $\FF_{\mathrm
a}(H_1\oplus H_2)$. Whether ${\frak H}_{\mathrm a}$ coincides with
\eqref{guhdesrw} or it is a proper subspace of it, is an open
problem (see also  Remark~\ref{zurftt} below). }\end{rem}

We next  define ${\frak H}_{\mathrm a,\,fin}{:=}{\frak H}_{\mathrm
a}\cap \FF_{\mathrm a,\,fin}(H\oplus H)$, ${\frak H}_{\mathrm
a,\,fin}$ being dense in ${\frak H}_{\mathrm a}$.  Let us consider
the $\rhoa(f)$'s as operators in ${\frak H}_{\mathrm a}$ with
domain ${\frak H}_{\mathrm a,\, fin}$.

\begin{lem}\label{ugtder} The operators $\rhoa(f)$\rom, $f\in
S(\R^d)$\rom, are essentially selfadjoint in ${\frak H}_{\mathrm a
}$\rom.\end{lem}

\noindent{\it Proof}. The operators are evidently symmetric. The
proof of  essential selfadjointness is quite standard (see e.g.\
\cite[Ch.~3, subsec.~3.8]{BeKo} and  \cite[Lemma~4.1]{Ly}), so we
only outline it.

As easily seen from the proof of Lemma~\ref{74dzt}, we have for
any $g^{(n)}\in \FF_{\mathrm a}^{(n)}(H\oplus H)$ $$
\rhoa(f)g^{(n)}=\sum_{j=1}^4 \rho_{{\mathrm a},\, j}(f)g^{(n)}\in
\bigoplus_{i=n-2,\,n,\,n+2}\FF_{\mathrm a}^{(i)}(H\oplus H),$$ and
moreover
\begin{equation}\label{zue53398}\|\rho_{{\mathrm a},\,j}(f)g^{(n)}\|_{\FF_{\mathrm a}(H\oplus
H)}\le C_1
\max\{\|f\|_{L^2(\R^d)},\|f\|_{L^1(\R^d)},\|f\|_{L^\infty(\R^d)}\}\,\|g^{(n)}\|_{\FF_{\mathrm
a }^{(n)}(H\oplus H)}\end{equation} for $j=1,\dots,4$ and some
$C_1>0$. From here, it is not hard to show that, for every
$g^{(n)}\in \FF_{\mathrm a}^{(n)}(H\oplus H)$, the series
$$\sum_{m=0}^\infty \frac {\|\rho_{{\mathrm a}}(f)^m
g^{(n)}\|_{\FF_{\mathrm a}(H\oplus H)}}{m!}\, t^m$$ converges for
$$ 0<t<\big(4C_1
\max\{\|f\|_{L^2(\R^d)},\|f\|_{L^1(\R^d)},\|f\|_{L^\infty(\R^d)}\}\big)^{-1}.$$
Therefore, any vector from ${\frak H}_{\mathrm a,\,fin}$ is
analytical for $\rhoa(f)$. By Nelson's analytic vector criterium
(e.g.\ \cite[Th.~X.39]{RS2}), the lemma follows.\quad
$\blacksquare$\vspace{2mm}

We denote by $\rhoa^\sim(f)$ the closure of $\rhoa(f)$ in ${\frak
H}_{\mathrm a}$, which is a selfadjoint operator by
Lemma~\ref{ugtder}.

\begin{lem}\label{he54w} For any $f_1,f_2\in S(\R^d)$, the
operators $\rho_{\mathrm a}^\sim(f_1)$ and $\rho_{\mathrm a
}^\sim(f_2)$ commute in the sense of their resolutions of the
identity\rom.\end{lem} \noindent {\it Proof}. Since $\rhoa(f_1)$
is essentially selfadjoint, the set $(\rhoa(f_1)+i\pmb1){\frak
H}_{\mathrm a,\, fin}$ is dense in ${\frak H}_{\mathrm a}$.
Furthermore, $(\rhoa(f_1)+i\pmb1){\frak H}_{\mathrm a,\,
fin}\subset {\frak H}_{\mathrm a,\, fin}$. Thus, by the proof of
Lemma~\ref{ugtder}, the operator
$\rhoa(f_1)\restriction(\rhoa(f_1)+i\pmb1){\frak H}_{\mathrm a,\,
fin}$ has a dense set of analytical vectors. Hence, the lemma
follows from \cite[Ch.~5, Th.~1.15]{BeKo}.\quad
$\blacksquare$\vspace{2mm}

\begin{th} \label{e5q2343} Let $k$ be the inverse Fourier
transform of a function $\hat k$ satisfying \eqref{rde4w3}\rom.
Let the Hilbert space ${\frak H}_{\mathrm a }$ and the operators
$\rhoa^\sim(f)$\rom, $f\in S(\R^d)$\rom, be defined as above\rom.
Then\rom, there exist a unique probability measure $\mu_{\mathrm
a}$ on $(S'(\R^d),{\cal B}(S'(\R^d)))$  \rom(${\cal B}(S'(\R^d))$
denoting the Borel $\sigma$-algebra on $S'(\R^d)$\rom) and a
unique unitary operator $I_{\mathrm a}:{\frak H}_{\mathrm a}\to
L^2(S'(\R^d),{\cal B}(S'(\R^d));\mu_{\mathrm a})$ such that
$I_{\mathrm a}\Omega=1$ and the following formula holds
\begin{equation}\label{ttre} I_{\mathrm a}\,\rhoa^\sim(f) I_{\mathrm a}^{-1}=\la \cdot,
f\ra\cdot,\qquad f\in S(\R^d).\end{equation}
\end{th}

\begin{rem}\rom{In terms of the spectral theory of commuting
selfadjoint operators (e.g.\ \cite{BeKo,Sam}),
Theorem~\ref{e5q2343} states that the family
$(\rhoa^\sim(f))_{f\in S(\R^d)}$ has a spectral measure
$\mu_{\mathrm a}$ on $(S'(\R^d),{\cal B}(S'(\R^d)))$. Furthermore,
since the operators $\rho_{\mathrm a}(f)$ have a Jacobi type form
in ${\cal F}_{\mathrm a}(H_1\oplus H_2)$, this result is close in
spirit to  \cite{bere,Ly}.}\end{rem}

\noindent {\it Proof of Theorem\/} \ref{e5q2343}. Let
$(h_k)_{k=0}^\infty$ be the sequence of Hermite functions forming
an orthonormal basis in $L^2(\R^d)$ and let $a_k>0$ be the
eigenvalue of the operator $A$ (defined by \eqref{z76t45})
belonging to the eigenvector $h_k$, $k\in\Z_+$.

We denote by $\R^\infty{:=}\R^{\Z_+}$ the space of all sequences
of the form ${\bf x}=(x_0,x_1,x_2,\dots)$, $x_k\in\R$, $k\in\Z_+$,
and we endow $\R^\infty$ with the product topology. The Borel
$\sigma$-algebra ${\cal B}(\R^\infty)$ coincides with the cylinder
$\sigma$-algebra ${\cal C}_\sigma(\R^\infty)$.

\begin{lem}\label{rtw342}   There exist a unique probability
measure $\tilde\mu_{\mathrm a}$ on $(\R^\infty,{\cal
B}(\R^\infty))$ and a unique unitary operator $\tilde I_{\mathrm a
}: {\frak H}_{\mathrm a}\to L^2(\R^\infty,{\cal B
}(\R^\infty);\tilde\mu_{\mathrm a})$ such that $\tilde I_{\mathrm
a }\Omega=1$ and\rom, for each $k\in\Z_+$\rom, $\tilde I_{\mathrm
a }\,\rhoa^{\sim}(h_k)\tilde I^{-1}_{\mathrm a}=x_k\cdot$\rom,
where $x_k\cdot$ denotes the operator of multiplication by
$x_k$\rom.
\end{lem}

\noindent{\it Proof}. For $f\in S(\R^d)$, we have
$f=\sum_{k=0}^\infty \langle f,h_k\rangle h_k$, where the series
converges in each space $S_p(\R^d)$, $p\in\N$, and hence in
$S(\R^d)$. Next, it follows from \eqref{zue53398} that, for each
fixed $G\in{\frak H}_{\mathrm a,\, fin}$, the mapping
\begin{equation}\label{4896743w} S(\R^d)\ni f\mapsto \rhoa(f)G\in{\frak
H}_{\mathrm a}\end{equation} is continuous. Therefore, $\Omega$ is
a cyclic vector for the family $(\rhoa^\sim(h_k))_{k=0}^\infty$.
Thus, $(\rhoa^\sim(h_k))_{k=0}^\infty$ is a countable family of
commuting selfadjoint operators having a cyclic vector, and hence
the lemma follows from \cite[Ch.~1, Th.~4]{Sam}.\quad
$\blacksquare$\vspace{2mm}

For each $p\in\N$, we define the following measurable function on
$\R^\infty$: $$\R^\infty\ni{\bf x}=(x_k)_{k=0}^\infty\mapsto
\|{\bf x}\|_{-p}^2{:=}\sum_{k=0}^\infty
x_k^2a_k^{-p}\in\R_+\cup\{+\infty\}.$$ Let $${\cal
S}_{-p}{:=}\{{\bf x}\in\R^\infty:\|x\|_{-p}^2<\infty\},\quad
p\in\N,\qquad {\cal S}'{:=}\bigcup_{p\in\N}{\cal S}_{-p}.$$
Evidently, ${\cal S}_{-p},\,{\cal S}'\in{\cal B}(\R^\infty)$. By
using the monotone convergence theorem and Lemma~\ref{rtw342}, we
get \begin{align} \int_{\R^\infty}\|{\bf
x}\|_{-p}^2\,d\tilde\mu_{\mathrm a
}(x)&=\int_{\R^\infty}\sum_{k=0}^\infty x_k^2 a_k^{-p}\,
d\tilde\mu_{\mathrm a}(x) \notag\\ &=\sum_{k=0}^\infty
a_k^{-p}\int_{\R^\infty} x_k^2\,d\tilde\mu_{\mathrm a}(x)
=\sum_{k=0} ^\infty a_k^{-p}\|\psi(h_k)\Omega\|_{{\frak
H}_{\mathrm a}}^2.\label{zur6z}\end{align} For some $C_2>0$,
    \begin{equation}\label{rtw348975}\max\{\|f\|_{L^2(\R^d)},\|f\|_{L^1(\R^d)},\|f\|_{L^\infty(\R^d)}\}\le
C_2\|f\|_{S_d(\R^d)},\qquad f\in S(\R^d),\end{equation} and since
the inclusion $S_d(\R^d)\hookrightarrow L^2(\R^d)$ is of
Hilbert--Schmidt type, \begin{equation}
\label{urt67}\sum_{k=0}^\infty a_k^{-d}<\infty.\end{equation} By
\eqref{zue53398}, \eqref{zur6z}--\eqref{urt67}, $$
\int_{\R^\infty}\|x\|_{-2d}^2\, d\tilde\mu_{\mathrm a}(x)\le C_1^2
C_2^2\sum_{k=0}^\infty a_k^{-d}<\infty.$$ This  yields that
\begin{equation}\label{zr564}\tilde\mu_{\mathrm a}({\cal S}_{-2d})=\tilde\mu_{\mathrm a}({\cal S}')=1.
\end{equation}
Let ${\cal B}({\cal S}')$ denote the trace $\sigma$-algebra of
${\cal B}(\R^\infty)$ on ${\cal S}'$. By \eqref{zr564}, we can
consider $\tilde\mu_{\mathrm a}$ as a probability measure on
$({\cal S}',{\cal B}({\cal S}'))$.

Noticing that  the mapping $${\cal S}'\ni{\bf
x}=(x_0,x_1,x_2,\dots)\mapsto {\cal E}{\bf x}{:=}\sum_{k=0}^\infty
x_kh_k \in S'(\R^d) $$  is a measurable bijection, we define a
probability measure $\mu_{\mathrm a}$ on $(S'(\R^d),{\cal
B}(S'(\R^d)))$ by $\mu_{\mathrm a}{:=}\tilde \mu_{\mathrm
a}\circ{\cal E}^{-1}$, and a unitary operator ${\cal U}:L^2({\cal
S}',{\cal B}({\cal S}');\tilde\mu_{\mathrm a})\to
L^2(S'(\R^d),{\cal B}(S'(\R^d));\mu_{\mathrm a})$ by $${\cal
U}F(\omega){:=}F({\cal E}^{-1}\omega),\qquad \omega\in S'(\R^d).$$
Setting $I_{\mathrm a}{:=}{\cal U}\tilde I_{\mathrm a}$, we get a
unitary operator acting from ${\frak H}_{\mathrm a}$ onto
$L^2(S'(\R^d);\mu_{\mathrm a})$ such that $I_{\mathrm a}\Omega=1$
and
\begin{equation}\label{hjre54w}I_{\mathrm a}\,\rhoa^\sim(h_k)
I_{\mathrm a}^{-1}=\la\cdot,h_k\ra\cdot,\qquad
k\in\Z_+.\end{equation} Furthermore, using the continuity of
mapping \eqref{4896743w}, we easily conclude from \eqref{hjre54w}
that \eqref{ttre}  holds.
 Thus, the theorem is proved.\quad $\blacksquare$

The configuration space $\Gamma_{\R^d}$ over $\R^d$ is defined as
the set of all locally finite subsets (configurations) in $\R^d$:
$$\Gamma_{\R^d}{:=}\big\{\,\gamma\subset\R^d\mid
|\gamma\cap\Lambda|<\infty\text{ for each compact
$\Lambda\subset\R^d$}\,\big\}.$$ Here, $|\Lambda|$ denotes the
cardinality of a set $\Lambda$. We can identify any
$\gamma\in\Gamma_{\R^d}$ with the positive Radon measure
$\sum_{x\in\gamma}\delta_x\in{\cal M}(\R^d)$, where $\delta_x$ is
the Dirac measure with mass at $x$,
$\sum_{x\in\varnothing}\delta_x{:=}$zero measure, and ${\cal
M}(\R^d)$ stands for the set of all positive Radon measures on
${\cal B}(\R^d)$. The space $\Gamma_ {\R^d}$ is endowed with the
relative topology as a subset of the space ${\cal M}(X)$ with the
vague topology. We denote by ${\cal B}(\Gamma_X)$ the Borel
$\sigma$-algebra on $\Gamma_{\R^d}$.

We endow ${\cal D}(\R^d)$ with its natural projective limit
topology and denote by ${\cal D}'(\R^d)$ the dual space of ${\cal
D}(\R^d)$. One can show that $\Gamma_{\R^d}$ belongs to the
cylinder $\sigma$-algebra ${\cal C}_\sigma({\cal D}'(\R^d))$, and
furthermore, the trace $\sigma$-algebra of ${\cal C}_\sigma({\cal
D}'(\R^d))$ on $\Gamma_{\R^d}$, resp.\ $S'(\R^d)$, coincides with
${\cal B}(\Gamma_X)$, resp.\  ${\cal B}(S'(\R^d))$. Thus, any
probability measure $\nu$ on $(S'(\R^d),{\cal B}(S'(\R^d)))$ can
be considered as a measure on $({\cal D}'(\R^d),{\cal
C}_\sigma({\cal D}'(\R^d)))$, and if additionally
$\nu(\Gamma_{\R^d})=1$, $\nu$ can be considered as probability
measure on $(\Gamma_{\R^d},{\cal B}(\Gamma_{\R^d}))$ as well.

Our next aim is to show that $\mu_{\mathrm a}$ is supported by
$\Gamma_{\R^d}$. To this end, let us recall the notion of
correlation functions of a probability measure $\nu$ on
$(\Gamma_{\R^d},{\cal B}(\Gamma_{\R^d}))$.

Let $\hot$ stand for the symmetric tensor product. For any
$g^{(n)}\in{\cal D}(\R^d)^{\hot n}$($=$the space of all smooth,
symmetric, compactly supported functions on $(\R^d)^n$), we define
a function $\Gamma_{\R^d}\ni\gamma\mapsto\la{:}\gamma^{\otimes
n}{:},g^{(n)}\ra\in\R$ by \begin{equation}\label{zudsfu}
\la{:}\gamma^{\otimes n}{:},g^{(n)}\ra{=}
\sum_{x_1\in\gamma}\,\sum_{x_2\in\gamma,\, x_2\ne x_1}\dots
\sum_{x_n\in\gamma,\, x_n\ne x_1,\dots,x_n\ne
x_{n-1}}g^{(n)}(x_1,\dots,x_n)\end{equation} (the number of the
non-zero summands on the right hand side of \eqref{zudsfu} is
finite). The  functions $(k_\nu^{(n)})_{n=1}^\infty$ with
$k_\nu^{(n)}:(\R^d)^n\to\R$ being measurable and symmetric, are
called correlation functions of the measure $\nu$ if, for each
$g^{(n)}\in{\cal D}(\R^d)^{\hot n}$, $n\in\N$,
\begin{equation}\label{hgfztdzzttz}
\int_{\Gamma_{\R^d}} \la{:}\gamma^{\otimes
n}{:},g^{(n)}\ra\,\nu(d\gamma)=\int_{(\R^d)^n}
g^{(n)}(x_1,\dots,x_n) k_{\nu}^{(n)}(x_1,\dots,x_n)\, dx_1\dotsm
dx_n\end{equation} (if the measure $\nu$ has correlation
functions, then these are a.s.\ uniquely defined).

As easily seen from \eqref{zudsfu}, the kernels
${:}\gamma^{\otimes n}{:}\in{\cal D}'(\R^d)^{\hot n}$ satisfy the
recursion relation
\begin{gather}{:}\gamma^{\otimes 1}(x){:}= \gamma(x),\notag\\  {:}\gamma^{\otimes
(n+1)}(x_1,\dots,x_{n+1}){:}=\big(\gamma(x_{n+1})\,{:}\gamma^{\otimes
n}(x_1,\dots,x_n){:}\notag\\\text{} -\sum_{i=1}^n
\delta(x_{n+1}-x_i)\,{:}\gamma^{\otimes
n}(x_1,\dots,x_n){:}\big)^\sim,\qquad n\in\N
,\label{hfztfdzw}\end{gather} where $(\cdot)^\sim$ denotes
symmetrization of a function. Replacing $\gamma\in\Gamma_{\R^d}$
with an arbitrary $\omega\in{\cal D}'(\R^d)$, we  may now  define
${:}\omega^{\otimes n}{:}\in{\cal D}'(\R^d)^{\hat\otimes n}$ and
introduce, analogously to \eqref{hgfztdzzttz}, the notion of
correlation functions $(k_\nu^{(n)})_{n=1}^\infty$ for any
probability measure $\nu$ on ${\cal D}'(\R^d)$. (We, however,
remark that the introduction of correlation functions for a
measure on ${\cal D}'(\R^ d)$ is only of ``technical'' nature,
since one always expects that a measure having correlation
functions is supported by $\Gamma_{\R^d}$, see the arguments
below).

Following \eqref{hfztfdzw}, we introduce operators
\begin{gather}\lw \rhoa(x)\rw=\rhoa(x),\notag\\
\lw\rhoa(x_{n+1})\rhoa(x_n)\dotsm
\rhoa(x_1)\rw=\big(\rhoa(x_{n+1})\,\lw
\rhoa(x_1)\dotsm\rhoa(x_n)\rw\notag\\ \text{}
-\sum_{i=1}^{n}\delta(x_{n+1}-x_i)\,\lw \rhoa(x_1)\dotsm
\rhoa(x_n)\rw\big)^\sim,\label{zur65e}\end{gather} which  make
sense after integration with test functions.

The following proposition shows that $\lw
\rhoa(x_n)\dotsm\rhoa(x_1)\rw$ is the ``normal product'' of the
the operators $\rhoa(x_1),\dots,\rhoa(x_n)$ (compare with
\cite[subsecs.~2.B and 2.C]{MSh}).

\begin{prop}\label{higztf} For each $n\in\N$ and $f_1,\dots,f_n\in S(\R^d)$\rom,
we have on ${\cal F}_{\mathrm a,\, fin}(H\oplus H)$\rom:
\begin{multline}\label{teeszr6} \int_{(\R^d)^n}dx_1\dotsm dx_n\,
f_1(x_1)\dotsm f_n(x_n) \,{:}\rhoa(x_1)\dotsm\rhoa(x_n)\rw \\
=\int_{(\R^d)^n} dx_1\dotsm
 dx_n\,f_1(x_1)\dotsm f_n(x_n)\,\psi^*(x_n)\dotsm\psi^*(x_1)\psi(x_1)\dotsm\psi(x_n).
\end{multline}\end{prop}

\noindent {\it Proof}. We first note that, for $n\ge2$, the
well-definedness of the operator on the right hand side of
\eqref{teeszr6} on ${\cal F}_{\mathrm a,\, fin}(H\oplus H)$  may
be proved by using arguments analogous to those as in the proof of
Lemma~\ref{74dzt}. We prove the proposition by induction. For
$n=1$, \eqref{teeszr6} is trivially satisfied. Suppose that
\eqref{teeszr6} holds for some $n\in\N$. Then, by
 the induction hypothesis, \eqref{tzew45}, and \eqref{zur65e},  we have
\begin{gather*}
\lw \rhoa(x_{n+1})\dotsm\rhoa(x_1)\rw
=\big(\psi^*(x_{n+1})\psi(x_{n+1})\psi^*(x_n)\dotsm\psi^*(x_1)\psi(x_1)\dotsm\psi(x_n)\\
\text{} -\sum_{i=1}^n
\delta(x_{n+1}-x_i)\psi^*(x_n)\dotsm\psi^*(x_1)\psi(x_1)\dotsm\psi(x_n)\big)^\sim\\
=\big(-\psi^*(x_{n+1})\psi^*(x_n)\psi(x_{n+1})\psi^*(x_{n-1})\dotsm
\psi^*(x_1)\psi(x_1)\dotsm \psi(x_n)\\ \text{}
-\sum_{i=1}^{n-1}\delta(x_{n+1}-x_i)\psi^*(x_n)\dotsm\psi^*(x_1)\psi(x_1)\dotsm\psi(x_n)\big)^\sim\\
=\big(\psi^*(x_{n+1})\psi^*(x_n)\psi^*(x_{n-1})\psi(x_{n+1})\psi^*(x_{n-2})\dotsm\psi^*(x_1)\psi(x_1)\dots\psi(x_n)\\
\text{}-\delta(x_{n+1}-x_{n-1})\psi(x_{n-1})^*\psi(x_n)^*\psi(x_{n-2})^*\dotsm
\psi^*(x_1)\psi(x_1)\dotsm \psi(x_n)\\ \text{}
-\sum_{i=1}^{n-1}\delta(x_{n+1}-x_i)\psi^*(x_n)\dotsm\psi^*(x_1)\psi(x_1)\dotsm\psi(x_n)\big)^\sim\\
=\big(\psi^*(x_{n+1})\psi^*(x_n)\psi^*(x_{n-1})\psi(x_{n+1})\psi^*(x_{n-2})\dotsm\psi^*(x_1)\psi(x_1)\dots\psi(x_n)\\
-\sum_{i=1}^{n-2}\delta(x_{n+1}-x_i)\psi^*(x_n)\dotsm\psi^*(x_1)\psi(x_1)\dotsm\psi(x_n)\big)^\sim\\
=\dots=\big((-1)^n\psi^*(x_{n+1})\dotsm\psi^*(x_1)\psi(x_{n+1})\psi(x_1)\dotsm\psi(x_n)\big)^\sim\\
=\psi^*(x_{n+1})\dotsm\psi^*(x_1)\psi(x_1)\dotsm\psi(x_{n+1}),
\end{gather*} the formulas above making sense after integration
with  test functions. \quad $\blacksquare$

\begin{prop}\label{jkgztd} For any $f_1,\dots,f_n\in S(\R^d)$\rom,
$n\in\N$\rom, \begin{multline}\label{z8t54}\bigg(
\int_{(\R^d)^n}dx_1\dots dx_n\, f_1(x_1)\dotsm f_n(x_n) \lw
\rhoa(x_1)\dotsm\rhoa(x_n)\rw\,\Omega,\Omega\bigg)_{{\frak
H}_{\mathrm a}}\\=\int_{(\R^d)^n}(f_1\hot\dotsm\hot
f_n)(x_1,\dots,x_n)\det(\kappa(x_i-x_j))_{i,j=1}^n\,dx_1\dotsm
dx_n,\end{multline}where $\kappa(x){:=}(2\pi)^{-d/2} k(x)$\rom,
$x\in\R^d$\rom.
\end{prop}
 \noindent{\it Proof}. By Proposition~\ref{higztf},
\begin{gather} \bigg( \int_{(\R^d)^n}dx_1\dotsm dx_n\,
f_1(x_1)\dotsm f_n(x_n) \lw
\rhoa(x_1)\dotsm\rhoa(x_n)\rw\,\Omega,\Omega\bigg)_{{\frak
H}_{\mathrm a}}\notag\\ =\int_{(\R^d)^n}(f_1\hot\dotsm\hot
f_n)(x_1,\dots,x_n)\big( a_1(\overline\varkappa_{1,x_n})\dotsm
a_1(\overline\varkappa_{1,x_1})a_1^*(\overline\varkappa_{1,x_1})\dotsm
a^*_1(\overline\varkappa_{1,x_n})\Omega,\Omega\big)_{{\frak
H}_{\mathrm a}}\, dx_1\dotsm dx_n \notag\\ =
\int_{(\R^d)^n}(f_1\hot\dotsm\hot
f_n)(x_1,\dots,x_n)\sqrt{n!}\,\big(
a_1(\overline\varkappa_{1,x_n})\dotsm
a_1(\overline\varkappa_{1,x_1})
 \overline\kappa_{1,x_1}\wedge\dotsm\wedge
\kappa_{1,x_n},\Omega\big)_{{\frak H}_{\mathrm a}}\, dx_1\dotsm
dx_n \notag\\   = \int_{(\R^d)^n}(f_1\hot\dotsm\hot
f_n)(x_1,\dots,x_n)\,n!\,\big( \overline\kappa_{1,x_1}\otimes
\dots \otimes\overline\kappa_{1,x_n},
\overline\kappa_{1,x_1}\wedge\dotsm\wedge\overline\kappa_{1,x_n}\big)_{{\frak
H}_{\mathrm a}}\, dx_1\dotsm dx_n \notag\\ =
\int_{(\R^d)^n}(f_1\hot\dotsm\hot f_n)(x_1,\dots,x_n) \det\big((
\overline\kappa_{1,x_i},\overline\kappa_{1,x_j})_{H_\C}\big)_{i,j=1}^n\,
dx_1\dotsm dx_n .\label{56rtzhb}\end{gather} Next, for any $f,g\in
S_\C(\R^d)$,
\begin{gather*} \int_{(\R^d)^2}(
\overline\kappa_{1,x},\overline\kappa_{1,y})_{H_\C}\,
f(x)\overline g(y)\, dx\, dy =
\int_{(\R^d)^3}\kappa_1(x-z)\overline\kappa_1(y-z)f(x)\overline
g(y)\, dx\, dy\, dz\\ = (K_1g,K_1f)_{H_\C}=(Kg,f)_{H_\C}
=\int_{\R^d}\int_{\R^d}(2\pi)^{-d/2}\overline{k(x-y)g(y)}\, dy\,
f(x)\, dx\\ = \int_{(\R^d)^2}\kappa(y-x)f(x)\overline g(y)\, dx\,
dy.\end{gather*} Hence,
\begin{equation}\label{jiuszutr}(
\overline\kappa_{1,x},\overline\kappa_{1,y})_{H}=\kappa(y-x)\qquad
\text{a.e.\ $(x,y)\in (\R^d)^2 $}.\end{equation} Furthermore, for
each $x\in\R^d$,
\begin{align} (
\overline\kappa_{1,x},\overline\kappa_{1,x})_{H_\C}&=\int_{\R^d}|\kappa_1(y-x)|^2\,
dy=\int_{\R^d}|\kappa_1(y)|^2\,dy=\int_{\R^d}(2\pi)^{-d}\hat
k(\lambda)\,d\lambda\notag\\ &=(2\pi)^{-d/2}(\FF^{-1}\hat
k)(0)=\kappa(0).\label{hjas}\end{align} Evidently, for any
$(x_1,\dots,x_n)\in(\R^d)^n$,
\begin{equation}\label{hghg}\det(\kappa(x_i-x_j))_{i,j=1}^n=\det(\kappa(x_j-x_i))_{i,j=1}^n.\end{equation}
Thus, \eqref{56rtzhb}--\eqref{hghg} imply \eqref{z8t54}.\quad
$\blacksquare$

\begin{cor} \label{zure54e} The measure $\mu_{\mathrm a}$ has correlation
functions, which are given by \begin{equation}\label{zutedrrd}
k_{\mu_{\mathrm a}}^{(n)}(x_1,\dots,x_n)=\det
(\kappa(x_i-x_j))_{i,j=1}^n,\qquad\text{\rom{$(x_1,\dots,x_n)\in(\R^d)^n$,
$n\in\N$}.}\end{equation}\end{cor}

\noindent {\it Proof}. By Theorem~\ref{e5q2343}, we have, for any
$f_1,\dots,f_n\in S(\R^d)$, $$ I_{\mathrm a}^{-1} \big( \la \lw
\cdot^{\otimes n}\rw,f_1\hot\dotsm\hot f_n\ra\big) =
\int_{(\R^d)^n}(f_1\hot \dotsm\hot f_n)(x_1,\dots,x_n) \lw
\rhoa(x_1)\dotsm \rhoa(x_n)\rw\,\Omega  .$$ From here and
Proposition~\ref{jkgztd} the statement easily follows. \quad
$\blacksquare$

\begin{th}\label{zfsders}
Let the conditions of Theorem~\rom{\ref{e5q2343}} be
fulfilled\rom. Then\rom, $\mu_{\mathrm a}(\Gamma_{\R^d})=1$\rom,
the correlation functions of $\mu_{\mathrm a}$ are given by
\eqref{zutedrrd}\rom, and the Fourier transform of $\mu_{\mathrm
a}$ is calculated as follows\rom: for each $f\in S(\R^d)$
\begin{equation}\label{zhgders98} \int
e^{i\la\omega,f\ra}\,\mu_{\mathrm
a}(d\omega)=\sum_{n=0}^\infty\frac1{n!}\int_{(\R^d)^n}(e^{if(x_1)}-1)\dotsm
(e^{if(x_n)}-1)\det(\kappa(x_i-x_j))_{i,j=1}^n\, dx_1\dotsm
dx_n.\end{equation}\end{th}

\noindent {\it Proof}.  We evidently have
$$|\kappa(x)|\le(2\pi)^{-d}\|\hat k\|_{L_1(\R^d)}\qquad \forall
x\in\R^d.$$ Hence, by \cite[Corollary~3]{M2} and
Corollary~\ref{zure54e}
\begin{equation}\label{hiwegfuw} |k_ {\mu_{\mathrm a}}^{(n)}(x_1,\dots,x_n)|\le
\big((2\pi)^{-d}\|\hat k\|_{L^1(\R^d)}\big)^n n^{n/2}\qquad
\forall (x_1,\dots,x_n)\in (\R^d)^n,\ n\in\N.\end{equation} By
\cite[Theorem~2]{BKKL}  (see also \cite[Theorem~6.5]{KK}), the
bound \eqref{hiwegfuw} implies that the measure $\mu_{\mathrm a}$
is concentrated on $\Gamma_{\R^d}$. Finally, formula
\eqref{zhgders98}  follows in a standard way  from
Lemma~\ref{zure54e} and bound \eqref{hiwegfuw} (see e.g.
\cite[Remark~2]{BKKL}).\quad $\blacksquare$\vspace{2mm}

By Theorem~\ref{zfsders}, $\mu_{\mathrm a}$ is a  fermion process
\cite{DV,Ma1}, or a determinantal random point field in terms of
\cite{Sosh}.

\begin{rem}\label{zurftt}\rom{ Let us suppose that, in addition to condition
\eqref{rde4w3}, the function $\hat k$ satisfies $$\int_{\R^d}\hat
k(\lambda)|\lambda|^n\,d\lambda<\infty,\qquad \forall n\in\N.$$
Then, using formula \eqref{u8zearg6}, for each $v\in {\cal
V}_0(\R^d)$, one can construct $J(v)$ as a selfadjoint operator in
${\cal F}(H_1\oplus H_2)$. Here, ${\cal V}_0(\R^d)$  denotes the
set of all smooth, compactly supported vector fields on $\R^d$.
Thus, one gets a representation of the full algebra $\frak g$ (see
Introduction). However, it is still an open problem, whether
${\frak H}_{\mathrm a}$ is an invariant subspace for the operators
$J(v)$. If it were so, we could evidently construct a
representation of the algebra $\frak g$, as well as the group $G$
in the space $L^2(\Gamma_{\R^d};\mu_{\mathrm a})$. }\end{rem}

\subsection{Bosonic case}

We now suppose that \eqref{763457z} holds. For each $x\in\R^d$, we
introduce a particle density operator $\rho_{\mathrm
s}(x){:=}\varphi^*(x)\varphi(x)$, acting continuously from
$\FF_{\mathrm s}(\Phi)$ into $\FF_{\mathrm s}^*(\Phi)$. Then, the
operators $$\rho_{\mathrm s}(f){:=} \int_{\R^d}dx\,
f(x)\rho_{\mathrm s}(x),\qquad f\in S(\R^d),$$ act continuously on
$\FF_{\mathrm fin}(H\oplus H)$.  Using \eqref{z7eawr76} and an
estimates of type \eqref{zue53398}, we show that $\rho_{\mathrm s
}(f)$ are essentially selfadjoint and their closures
$\rho^\sim_{\mathrm s}(f)$ constitute a cyclic family of commuting
selfadjoint operators in the Hilbert space ${\frak H}_{\mathrm
s}$---the closure of the linear span of the vectors $$
\big\{\,\Omega,\, \rho_{\mathrm s }(f_1)\dotsm\rho_{\mathrm
s}(f_n)\Omega\mid f_1,\dots,f_n\in S(\R^d),\ n\in\N  \,\big\}$$ in
$\FF_{\mathrm s}(H\oplus H)$.

We then construct the spectral measure $\mu_{\mathrm s}$ of the
operator family $(\rho^\sim_{\mathrm s}(f))_{f\in S(\R^d)}$ as a
probability measure on $(S'(\R^d),{\cal B}(S'(\R^d)))$.
Furthermore, with the help of formula \eqref{gtfr56e54} we show
that $\mu_{\mathrm s}$ has correlation functions, which are given
by the following formula:
\begin{equation}\label{zudrrd}
k_{\mu_{\mathrm s}}^{(n)}(x_1,\dots,x_n)=\per
(\kappa(x_i-x_j))_{i,j=1}^n,\qquad\text{\rom{$(x_1,\dots,x_n)\in(\R^d)^n$,
$n\in\N$},}\end{equation} where $\kappa(x){:=}(2\pi)^{-d/2}k(x)$.
Next, for every bounded $\Lambda\in{\cal B}(\R^d)$, we evidently
have the following estimate:
\begin{equation}\label{zure64e} \frac1{n!}\int_{\Lambda^n}|k_{\mu_{\mathrm s}}^{(n)}(x_1,\dots,x_n)|
\, dx_1\dotsm dx_n\le \big(|\Lambda|\,C_3\big)^n,\end{equation}
where $|\Lambda|$ denotes the volume of $\Lambda$ and
$C_3{:=}\sup_{x\in\R^d}|\kappa(x)|<\infty$. Hence, by
\eqref{zure64e} and \cite[Th.~2]{BKKL}, we get $\mu_{\mathrm s
}(\Gamma_{\R^d})=1$. Thus, we get the following

\begin{th} Let $k$ be the inverse Fourier transform of a function
$\hat k$ satisfying \eqref{763457z}\rom. Let the Hilbert space
${\frak H}_{\mathrm s}$ and the operators $\rho^\sim_{\mathrm
s}(f)$\rom, $f\in S(\R^d)$\rom, be defined as above\rom. Then\rom,
there exist a unique probability measure $\mu_{\mathrm s}$ on
$(S'(\R^d),{\cal B }(S'(\R^d)))$ and a unique unitary operator
$I_{\mathrm s}:{\frak H}_{\mathrm s}\to L^2(S'(\R^d);\mu_{\mathrm
s})$ such that $I_{\mathrm s}\Omega=1$ and $$I_{\mathrm
s}\,\rho^\sim_{\mathrm s}(f)I_{\mathrm s}^{-1}=\la
\cdot,f\ra\cdot,\qquad f\in S(\R^d).$$ Furthermore\rom,
$\mu_{\mathrm s}(\Gamma_{\R^d})=1$ and the correlation functions
$(k_{\mu_{\mathrm s}}^{(n)})_{n=0}^\infty$ of the measure
$\mu_{\mathrm s}$ are given by formula \eqref{zudrrd}\rom.\end{th}

By \cite{Ma0, Ma1} (see also \cite{DV}), $\mu_{\mathrm s}$ is a
 boson process.\vspace{2mm}

\noindent {\bf Acknowledgements}

I am  grateful to Yu.~Kondratiev for drawing my attention to the
fermion processes and for his permanent interest in this work. I
would like to thank S.~Albeverio, G.~Goldin and Yu.~Samoilenko for
useful discussions. I am also grateful to the referees of the
paper for many suggestions on improvement of the first version of
the paper. The financial support of SFB 256, DFG Research Projects
436 RUS 113/593, and BMBF Research Project UKR-004-99 is
gratefully acknowledged.

\end{document}